\pdfoutput=1
\documentclass[final,3p,times,twocolumn]{elsarticle}

\usepackage{amssymb}
\usepackage{eurosym}
\usepackage{multirow}
\usepackage[table]{xcolor}
\usepackage[hyphens]{url}

\usepackage{graphicx}
\usepackage{epstopdf} %converting to PDF

\definecolor{beaublue}{rgb}{0.74, 0.83, 0.9}
\definecolor{lightgrey}{rgb}{0.94, 0.94, 1.0} % a little darker for the tables

\usepackage{lineno}

\journal{Energy Policy}

\begin{document}

\begin{frontmatter}

\title{Estimating the Impact of Wind Generation in the UK}

\author{Lisa M.H. Hall}
\ead{lisa.clark@sheffield.ac.uk}

\author{Alastair R. Buckley}
\author{Jose Mawyin}

\address{Department of Physics and Astronomy, Hounsfield Road, Sheffield S3 7RH, United Kingdom}

\begin{abstract}
This paper studies the impact of wind generation on market prices and system costs in the UK between 2013 and 2014.  The wider effects and implications of wind generation is of direct relevance and importance to policy makers, as well as the grid operator and market traders.  We compare electricity generation from Coal, Gas and wind, on both the wholesale and imbalance market.  We calculate the system cost of wind generation (government subsidies and curtailment costs) and the total energy costs.  For the first time in the UK, we calculate the Merit Order Effect on spot price due to the wind component and show a $1.67\%$ price decrease for every percentage point of wind generation (compared to the ``zero-wind'' price).  The net result of total costs and price savings is roughly zero (slight positive gain).  We also consider the effect of not having either an onshore or an offshore wind component.  We show that the Merit-Order Effect savings are heavily reduced, leading to an outgoing cost of wind generation in both cases.  It is therefore important to have a significant total percentage of wind generation, from both onshore and offshore farms.
\end{abstract}


\end{frontmatter}

%%\linenumbers


\section{Introduction} % (fold)
\label{sec:introduction}

Investment in energy systems tends to be as much political as it is technological or scientific.  Despite calls from the international community to retain investment mechanisms for onshore wind in the UK (because on-shore wind is by far the cheapest renewable generation), in May 2015, the newly-elected government quickly removed support, citing over-expensive subsidy costs.  Additionally, in January 2015, several UK broadsheets led with stories claiming that wind farms are over-charging the market to curtail generation during low-demand periods.  It is notable that both subsidies and curtailment costs are typically not discussed in comparison with fossil fuel electricity costs, nor in terms of the financial benefits to the market.

Whilst renewable subsidies might seem costly, the expected outcome of subsidising such technologies in the market is two-fold: to decarbonise future energy supply and reduce the price of energy in the long-term.  Discussing the cost of both subsidies and curtailment costs without defining the context can be misleading and can lead to the wrong conclusion.

In Germany, it is widely accepted that a supply of electricity from renewable generators (predominately wind and solar PV) reduces the spot price and leads to a considerable cost saving.  Several studies have calculated this Merit-Order Effect (MOE), though analyses vary.  For example, Senfu\ss~et al show that the MOE in 2006 led to a price reduction of about $8$ \euro/MWh and total costs saving of about \euro5 billion  in 2006\cite{Sensfuss20083086}.  Whilst they cite enormous support payments around \euro 5.6 billion, they report a net profit due to the value of renewable electricity and the MOE cost saving.  

Similarly, Weigt calculated savings in Germany of \euro 1.3 billion in 2006, \euro 1.5 billion in 2007, and \euro 1.3 billion in the first half of 2008\cite{Weigt20091857}.  Ketterer shows a $1.46\%$ spot price drop as wind increases by 1 percentage point\cite{Ketterer2014270}, whilst Cludius et al show a spot market price reduction of 6 \euro/MWh in 2010 rising to 10 \euro/MWh in 2012\cite{Cludius2014302} due to a combined electricity generation by wind and PV.  Similar results can be shown for the Spanish market\cite{SaenzdeMiera20083345} and Irish market\cite{Swinand2015468}, though the effect is heavily reduced in Ireland.

For the UK market, comparable analyses have not been undertaken.  Indeed, the MOE has not been quantified, nor has the comparative cost of wind curtailment.  The reason for this may lie in the fact that wind generation has only been a significant factor for a few years and until now, such analysis would be meaningless.  Additionally, analysis of the electricity market in the UK is impeded by several issues: access to market data, usability of accessible data and the complexity of the market structure. 

In this paper, for the first time, we calculate the value of the Merit Order Effect in the UK and compare it to subsidy costs.   Furthermore, we present the cost of curtailment of both renewable and conventional generation (coal and gas) in the UK power markets.  Where possible, we differentiate between onshore and offshore wind capacity, as well as a geographic split (England and Scotland, according to operating capacity during the relevant timescales).

It is interesting to note that other researchers are now considering the implications of curtailing versus storing renewable energy\cite{Barnhart2013}, but since we do not consider instantaneous curtailment costs, we do not comment on the wider consequences for the UK market.

% section introduction (end)

% figure of merit order
% placed here to position it within the paper
\begin{figure}[tb!]
\centering
\includegraphics[width=\columnwidth]{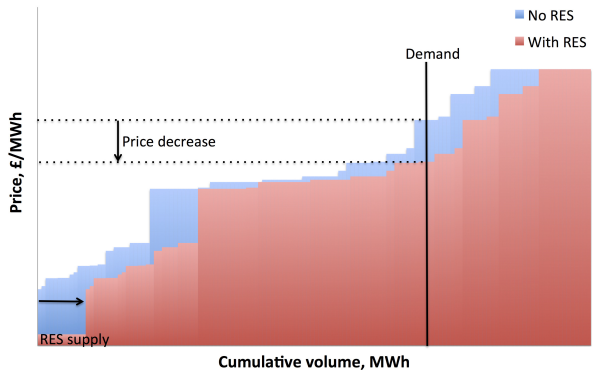}
\caption{The Merit Order Effect (MOE).  An increased, cheap supply of renewable energy pushes the curve to the right (i.e. the blue curve becomes the red curve) and results in a decreased marginal cost of energy per unit.} 
\label{fig:merit_order}
\end{figure}

\section{UK Spot Market and Balancing Mechanism} % (fold)
\label{sec:uk_wholesale_and_imbalance_market}

Since electricity cannot be stored in large amounts, supply and demand must be balanced at all times.  In the UK, this is undertaken by a system of trading on the wholesale market by generators, suppliers and customers.  Electricity can also be imported or exported through international interconnectors.  Trades are completed for each half-hour period, called Settlement Periods (SPs).  All trading must be completed one hour before the relevant half-hour period (called gate closure), at which time the grid operator (National Grid Electricity Transmission) must balance the supply and demand to ensure instantaneous match.  Typically, the Balancing Mechanism (BM) makes up around 2-5\% of total electricity trades. 

The value of electricity on the wholesale market can be quantified using the electricity spot price, which is calculated as a volume-weighted average of all trades executed through the UK APX Power exchange.  The spot-price is calculated for each Settlement Period, but does not contain any information regarding fuel type.  At gate closure, all Balancing Mechanism Units (BMUs = generators and suppliers) notify the grid operator of their dispatch profile and also submit \emph{bids} and \emph{offers} of prices to deviate from their contracted generation/supply.  An offer is a proposal to increase generation or reduce demand.  A bid is a proposal to reduce generation or increase demand.

The grid operator follows the Balancing and Settlement Code (BSC) in order to allow the grid to be balanced at the best price available to the market.  Essentially, the bids and offers are tallied against the imbalance volume and appropriate offers/bids are accepted.  The System Buy Price (SBP) and System Sell Price (SSP) are calculated according to the most expensive 500MWh of energy supplied to the imbalance market (volume-weighted average price). Elexon provides a useful guidance document to the imbalance pricing, which details the exact calculation\cite{Routier2014}.

Within the balancing mechanism, offers are always positive (it will cost the system to increase generation), but bids can be be either positive or negative (generators may pay to reduce their generation, but may decide to charge the system for curtailing their generation).  With the increase of renewable electricity generation, negative bidding has become more prevalent as the generators attempt to counteract lost subsidies.  When the grid operator is required to curtail wind generation in order to balance the market, such negatively priced bids result in a cost to the system and the SSP may also become negative.  

On the other side, an increase in renewable energy supply leads to a lower system price, due to the Merit Order Effect.  This effect is shown pictorially in Figure~\ref{fig:merit_order}.  An increased, cheap supply of renewable energy pushes the supply curve to the right and results in a decreased volume-weighted average cost of energy.  This is true for the wholesale market, but the effect in the balancing market is reduced due to negative bidding.  In order to quantify the net gain or loss, it is imperative to consider both costs and savings, which has not been done previously in the UK.

% section uk_wholesale_and_imbalance_market (end)

% input table of total annual generation (main bulk and imbalance)
% put here to position it within the paper
\begin{table*}[t]
\centering
\small
\begin{tabular}{|l|c|c|c|c|c|c|c|c|}
\hline
\multirow{3}{*}{Fuel Type} & \multicolumn{2}{c|}{Total Generation, TWh} & \multicolumn{6}{c|}{Imbalance Volume as percentage of Total, \%}     \\ \cline{2-9} 
                           & \multirow{2}{*}{2013}    & \multirow{2}{*}{2014}   & \multicolumn{3}{c|}{2013}              & \multicolumn{3}{c|}{2014}              \\ \cline{4-9} 
                           &                          &                         & Offers & Positive bids & Negative bids & Offers & Positive bids & Negative bids \\ \hline

CCGT  &  216.88  &  236.07  &  3.047  &  -2.004  &  -0.002  &  2.461  &  -1.616  &  0.000\\ 
COAL  &  276.40  &  224.36  &  0.361  &  -1.311  &  -0.051  &  0.550  &  -1.709  &  -0.101\\ 
Off. Wind (Eng.)  &  24.76  &  24.76  &  0.001  &  0.000  &  -3.823  &  0.001  &  0.000  &  -0.046\\ 
Off. Wind (Scot.)  &  0.61  &  3.46  &  0.000  &  0.000  &  0.000  &  0.001  &  0.000  &  -0.005\\ 
On. Wind (Eng.)  &  1.83  &  2.46  &  0.000  &  0.000  &  0.000  &  0.000  &  0.000  &  0.000\\ 
On. Wind (Scot.)  &  17.93  &  20.51  &  2.911  &  -0.960  &  -0.001  &  1.988  &  -0.059  &  -0.032\\ \hline
All fuels  &  717.32  &  690.66  &  1.182  &  -1.373  &  -0.029  &  1.264  &  -1.375  &  -0.046\\ 
All Wind  &  45.13  &  51.20  &  1.157  &  -0.436  &  -0.027  &  0.797  &  -0.598  &  -0.035\\ \hline

\end{tabular}
\caption{Breakdown of the annual electricity generation by fuel type.  Energy generation is separated into total generation (left-hand columns) and the imbalance market (right-hand columns). Wind generation is split into onshore generation (labelled ``On.'' for brevity) and offshore generation (labelled ``Off.'').  Wind is also split into Scottish and English locality (labelled ``Scot.'' and ``Eng.'') respectively.  It can be seen that coal and gas are curtailed by around 1-3\% with predominantly positive bids (cost to supplier), whereas wind is curtailed around 0.5\% with both positive and negative bids (cost to both supplier and grid operator). Source: BMReports}
\label{tab:generation}
\end{table*}

\section{Data and Methodology} % (fold)
\label{sec:data}

We have collated electricity generation data from various sources.  Whilst we detail specific data sources below, a qualitative overview of the methodology will first be given.  Using electricity generation data for each half-hourly period between August 2012 and May 2015, it is possible to separate the two markets: the wholesale market (with spot price) and the imbalance market (with system buy and sell prices).  Using half-hourly data from the wholesale market, it is possible to correlate the spot price to the total percentage of wind generation, whilst the imbalance market is useful to understand the curtailment costs of generation by fuel type (i.e. the cost of turning off specific generators).  The data sources we utilise give generation volumes according to each generating unit across the UK and we associate these units to their associated fuel type.  We can therefore identify half-hourly generation in both the wholesale and imbalance markets according to fuel type.  This allows us to separate not only wind from fossil-fuel generation, but we can also distinguish onshore and offshore wind generation.  Due to the nature of the data, complications arise from matching and joining data sources (by both settlement period and generating unit), though statistical analysis of the resulting dataset is straightforward.  For example, one complication lies within the definition of date and time; some sources use local time, some give time in GMT and others only give Settlement Period.

We now detail the exact data sources and specific methodology.  Spot price data was obtained by academic license from APX Group and is also available per settlement period.  Wholesale electricity generation disaggregated by fuel type is available from two separate data providers, originating from the sources: (a) the Elexon Portal publishes data for each settlement period, bundled in quarter-yearly periods and (b) the BMReports website publishes the Final Physical Notification (FPN) data, which can be linked to fuel type by individual generation plant information.  Information regarding the Balancing Mechanism was sourced from the BMReports website, using the system prices (SYSPRICE data) and detailed system prices (DETSYSPRRICE data).  It should be noted that the FPN and Balancing Mechanism data is published online (to BMReports) shortly after each settlement period and is never updated.  However, the information at time of publication is not always complete (some transactions occur post-publication).  The complete (and therefore more correct) data is maintained by Elexon, though they do not widely publish as much disaggregated data.  The difference between data is minimal and should not alter any overarching results.  However, it is important to acknowledge that there is a compromise between the use of (slightly) incomplete data sets and breakdown of information.  For example, the Elexon data does not distinguish between onshore or offshore wind generation, so we must rely on the incomplete BMReports data.

The fuel types categories are not standardised between the datasets, so we limit our analysis to just CCGT (Combined Cycle Gas Turbine), Coal and Wind, which are consistent across the sets.  We have chosen CCGT and Coal as they are the predominant fossil fuel generators.  The date availability of data also differed between the datasets.  For most of our analysis, we therefore consider timescales for which we have a complete annual data (two full years: 2013 and 2014), but extend the timescale when we analyse the Merit Order Effect (from August 2012 until March 2015).  

We have correlated the three datasets and analysed them statistically using the SAS JMP program.  When considering the mean of a dataset, we always calculate the volume-weighted mean, such that insubstantial volumes do not skew the analysis.  Additionally, when discussing the balancing mechanism data, it is imperative to split the offers and bids into three (not two) separate categories:  offers, positive bids and negative bids.  This is undertaken to ensure that cancelling effects do not occur. 

It will be necessary later to use the Transmission Loss Multiplier (TLM).  However, the BMReports data associated with each offer/bid is not complete with respect to the these values (if the offer/bid is not part of the SSP/SBP price calculation, then the TLM is not published).  Equally, the BMReports publish a separate historic TLM dataset, which does not fully correspond to the offer/bid TLM data.  Furthermore, the TLM value depends on whether the BMU is a producer or consumer and this data is not readily available (we have access to only an incomplete list for all BMUs).  Notably, a producer has a ``delivering'' TLM (less than unity) and a consumer has an ``off-taking'' TLM (greater than unity).  It is therefore necessary to utilise data from three datasets (BMReports historic TLM, BMReports DETSYSPRICE data and Elexon historic TLM) by following a given procedure:
\begin{itemize}
	\item If the BMU is a known producer/consumer, then:
		\begin{itemize}
			\item If the Elexon TLM data exists for that SP, use the Elexon TLM data.
			\item If the Elexon TLM data does not exist for that SP, use the BMReports historic TLM dataset.
		\end{itemize}
	\item If it is not known whether the BMU is a producer or consumer, utilise the published DETSYSPRICE TLM data value:
		\begin{itemize}
			\item If the DETSYSPRICE TLM value is $<1$, then assume the BMU is a producer (and use the above rules)
			\item If the DETSYSPRICE TLM value is $>1$, then assume the BMU is a consumer (and use the above rules)
		\end{itemize}
\end{itemize}
This procedure defines all the offers/bids within our dataset.
% section data (end)
 
For calculation of the wind subsidies we utilise data published by the UK energy regulator, Ofgem, on the Renewable Obligation Certificates (ROCs) and Feed-In Tariff (FIT) scheme.
 
\section{Results} % (fold)
\label{sec:results}

% graph of cash flow
% put here to position within the paper
\begin{figure}[tb!]
\centering
\includegraphics[width=\columnwidth]{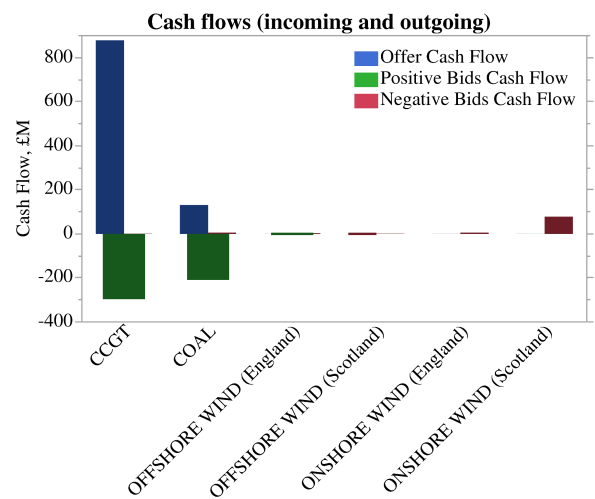}
\caption{The imbalance market: the cash flow summation of offers and bids (positive and negative) for 2013 and 2014.  Cash flow is taking from the perspective of the grid operator (positive flow is paid out, negative flow is cash income). Source: BMReports} 
\label{fig:Incoming_Outgoing}
\end{figure}

The total energy generation statistics are shown in Table~\ref{tab:generation}, as well as the breakdown into CCGT, Coal and Wind (onshore and offshore, split into geographic regions), for the years 2013 and 2014.  Alongside these numbers, we have calculated the corresponding imbalance volumes as a percentage of the total generation.  It can be seen that the imbalance volume typically makes up 2-3\% of the total generation volume.  The imbalance volume has been split into offers and bids, which shows that CCGT and Coal make up the majority of the offers and positive bids, whilst wind generation accounts for much of the negative bid volume (curtailment).  

Within public dissemination articles (media and public engagement), the high curtailment costs of wind generation has been commonly used as an argument against wind subsidies and renewable technology.  The main argument is that wind generators often submit very significant negative bids when compared to the other generators and are therefore expensive to curtail.  We will now explore this statement.

The breakdown of the cash flow associated with the offers and bids detailed in Table~\ref{tab:generation} is shown in Figure~\ref{fig:Incoming_Outgoing}.  The cash flow for each component ($i=\textrm{offers}, \textrm{positive bids}, \textrm{negative bids}$) is calculated as:
\begin{equation}
\textrm{Cash flow}_i	= \sum_j V * P *\textrm{TLM}
\end{equation}
where $j$ corresponds to the set of settlement periods, $V$ is the bid volume, $P$ is the offer/bid price and $TLM$ is the associated Transmission Loss Multiplier. 

\begin{figure}[tb!]
\centering
\includegraphics[width=\columnwidth]{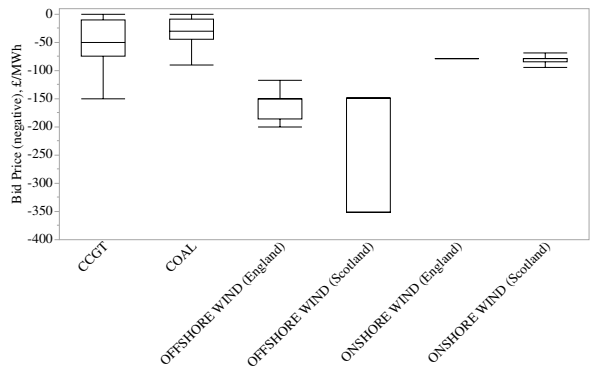}
\caption{The breakdown of negative bid prices for CCGT, COAL and Wind generation in 2013 and 2014. The horizontal line within each box represents the median sample value, the ends of the box represent the 25th and 75th quartiles and the whiskers (extending from each box) are determined by 1st(3rd) quartile -(+) interquartile range.  Source: BMReports} 
\label{fig:Negative_Bids}
\end{figure}

\begin{figure}[tb!]
\centering
\includegraphics[width=\columnwidth]{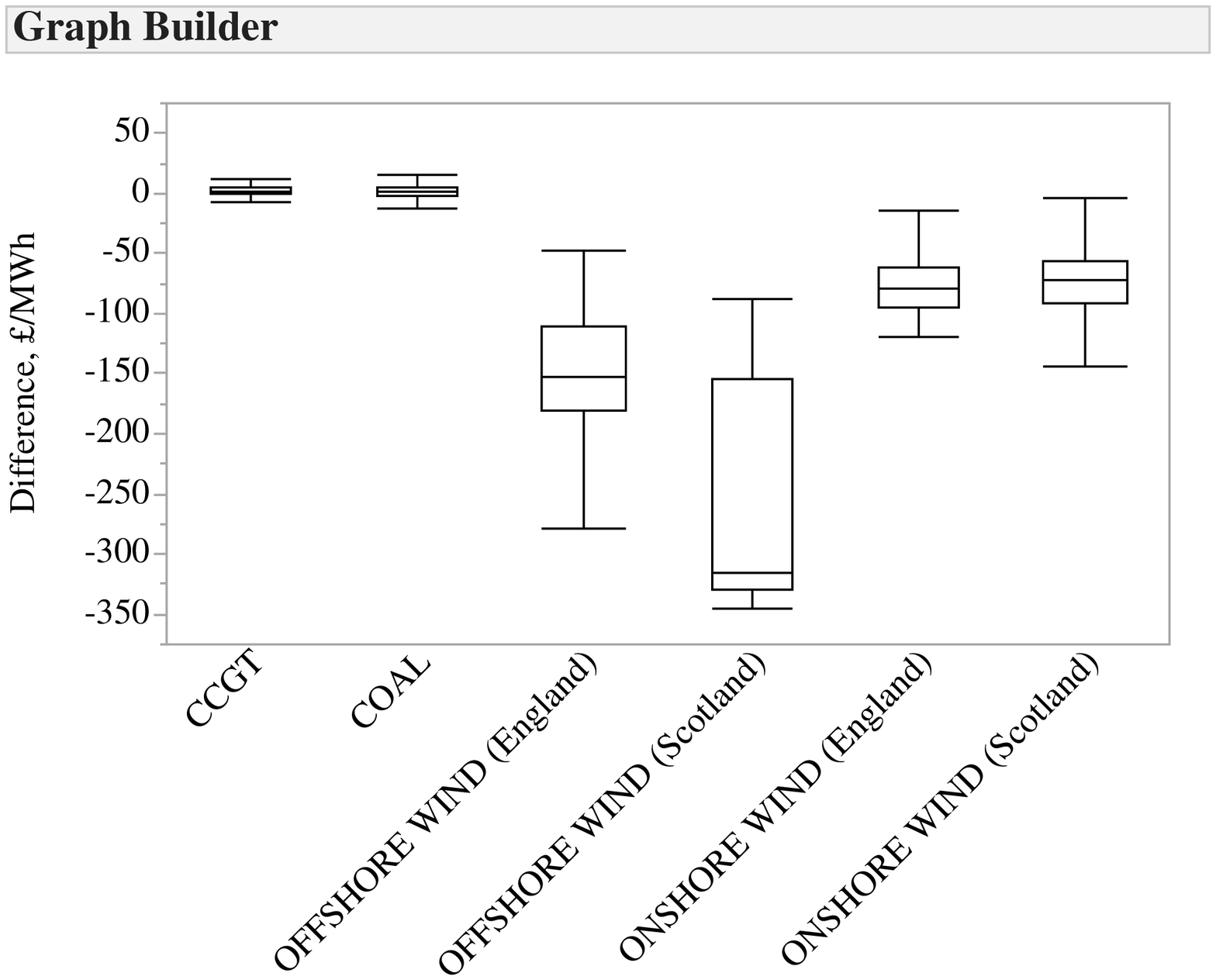}
\caption{The variance of difference between the bid price (per bid) and volume-weighted mean bid price (per settlement period) within 2013 and 2014. The horizontal line within each box represents the median sample value, the ends of the box represent the 25th and 75th quartiles and the whiskers (extending from each box) are determined by 1st(3rd) quartile -(+) interquartile range.  Bids from wind generators tend to be heavily below the volume-weighted mean bid price within a settlement period. Source: BMReports} 
\label{fig:Diff_from_volume-weighted_bids}
\end{figure}

We now consider just the set of negative bids.  From Figure~\ref{fig:Incoming_Outgoing} it can be seen that Scottish onshore wind generators comprise the huge majority of cash flow relating to negative bids (nearly \pounds90M) over the two year period (2013-2014).  The negative bids cash flow can be further broken down into the variance of negative bids (shown in Figure~\ref{fig:Negative_Bids}) and the difference between the bid price (per bid) and the volume-weighted mean bid price (per settlement period) (shown in Figure~\ref{fig:Diff_from_volume-weighted_bids}).  The latter statistic is a measure of how close (or far away) each single negative bid is from the corresponding bids during the same settlement period.  As can be seen (Figure~\ref{fig:Incoming_Outgoing}), the mean wind negative bid prices well exceed the lost subsidy of 55 \pounds/MWh (i.e. negative bid price significantly lower than -55 \pounds/MWh).  Additionally, whilst CCGT and Coal negative bids roughly correlate with the mean Settlement Period bid price (consistent with a zero difference), the wind bids are statistically more negative by a significant level (Figure~\ref{fig:Diff_from_volume-weighted_bids}).

At this point, it is also worth considering the level of subsidies paid to wind generators.  There are several subsidy policies available in the UK, which include the Renewable Obligation Certificates (ROCs) and Feed-In Tariffs (FITs).  In this simple analysis, we assume that the significant majority of the subsidies arise from the ROCs. The number of ROCs and the final cost (broken down into obligation periods between 2012 and 2015) is shown in Table~\ref{tab:ROCs}.  Roughly speaking, the annual wind subsidies lie around \pounds1.8 billion.  We compare this to the estimated cost of the FIT scheme in 2013-2014 of around \pounds97M (\pounds56M in 2012-2013 and only \pounds6M in 2011-2012).
% input table of total actual costs and predicted low wind costs
\begin{table*}[t]
\centering
\begin{tabular}{|l|ccc|ccc|}
\hline
\multirow{2}{*}{} & \multicolumn{3}{c|}{Number of ROCs (millions)} & \multicolumn{3}{c|}{Cost of ROCs (£bn)} \\ \cline{2-7} 
                  & 2012-13        & 2013-14       & 2014-15       & 2012-13  	 & 2013-14     & 2014-15    \\ \hline
Offshore wind     & 15.69          & 23.94         & 25.37         & 0.639    & 1.006   & 1.099   \\ 
Onshore wind      & 12.21          & 18.71         & 17.73         & 0.497    & 0.786    & 0.768    \\ \hline
TOTAL             & 27.9           & 42.65         & 43.1          & 1.136    & 1.792    & 1.866    \\ \hline
\end{tabular}
\caption{The number of Renewable Obligation Certificates (ROCs) issued per obligation period (April to March) between 2012 and 2015 and the associated level of costs. Source: Ofgem}
\label{tab:ROCs}
\end{table*}

\begin{figure}[tb!]
\centering
\includegraphics[width=\columnwidth]{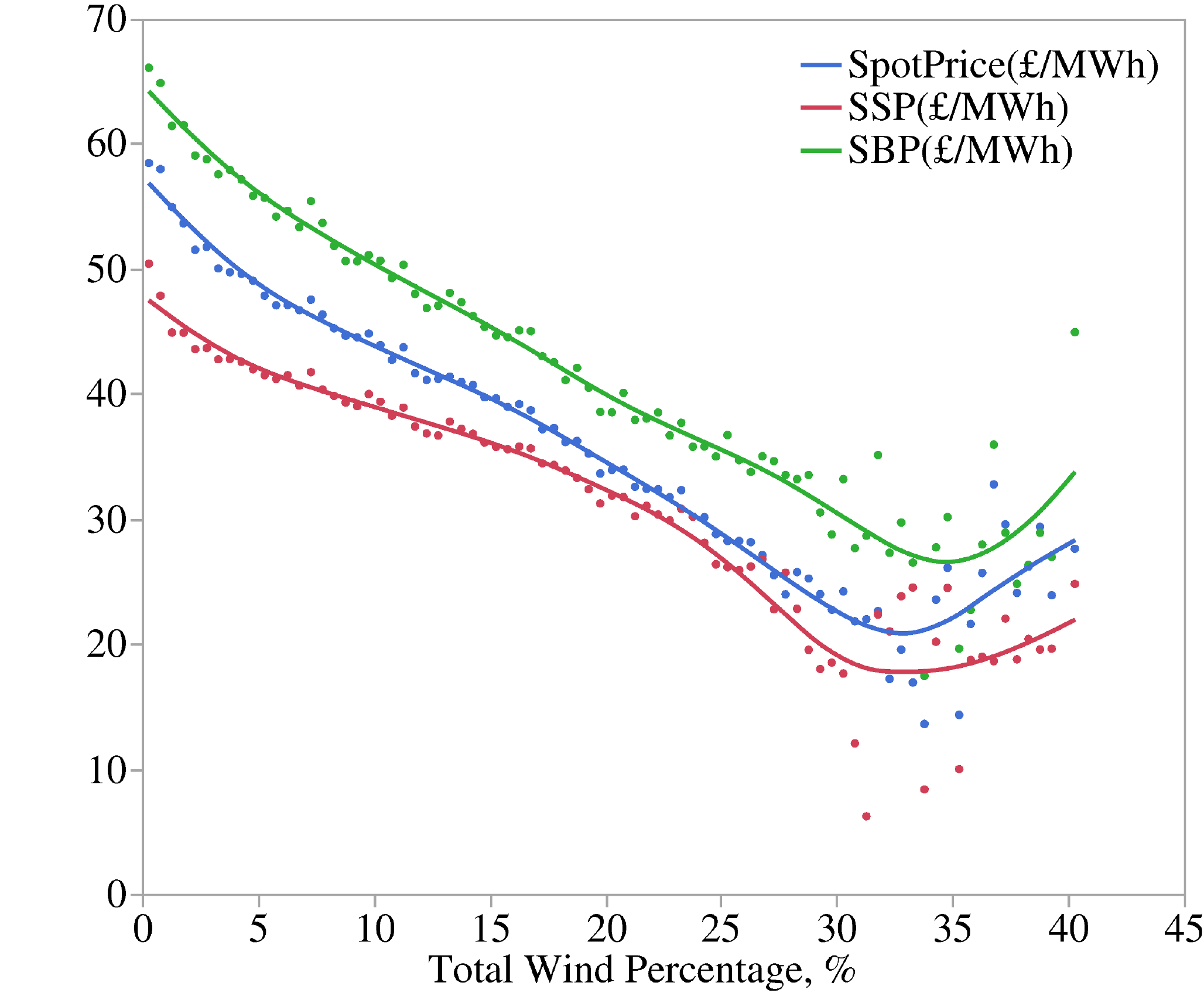}
\caption{The effect of increasing wind generation (as percentage of energy mix) on System Sell Price (SSP), System Buy Price (SBP) and Spot Price.  Linear fits are quoted in the table above and below the ``knee'', where a fit is assumed to be $y=y_0+mx$ and $x_0$ is the x-axis intercept.  Data is taken from August 2012 until March 2015.  Source: APX group and BMReports} 
\label{fig:prices_vs_wind}
\end{figure}

% input table of line of fit for wind pricing
\begin{table}[ht]
\centering
\scriptsize
\begin{tabular}{|l|ccc|}
\hline
     & $x_0$     & $y_0$     & m      \\ \hline
SSP  &  59.34  &  46.88  &  -0.79 \\
SBP  &  64.62  &  60.10  &  -0.93 \\
Spot Price  &  60.06  &  52.85  &  -0.88 \\ \hline

\end{tabular}
\caption{Best linear fit for Merit Order Effect (up to 25\%) in Figure~\ref{fig:prices_vs_wind}.  Linear fit assumed to be $y=y_0+mx$ and $x_0$ is the x-axis intercept. $m$ has units of \pounds/MWh/\%.}
\label{tab:gradients}
\end{table}

% figure of correlation of Onshore/Offshore wind
% placed here to position it within the paper
\begin{figure}[tb!]
\centering
\includegraphics[width=\columnwidth]{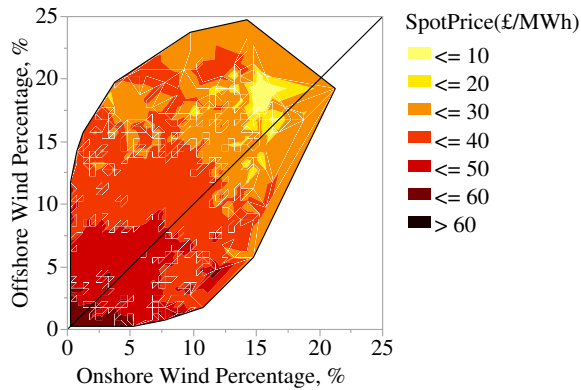}
\caption{The correlation between onshore and offshore wind generation (by percentage of total generation) to Spot Price.  The colour relates to the right-hand colorbar and represents the mean spot price (£/MWh).  The black line around the contour delineates the location of data points (i.e. all data lies within this outer edge).  The diagonal black line represents the equality of onshore and offshore wind percentages.  The majority of data points lie slightly above this line, which shows that offshore wind generation is typically higher than that for onshore wind.  Data is taken from August 2012 until March 2015.   Source: APX group and BMReports} 
\label{fig:wind_price_contour}
\end{figure}

The next point to consider is the Merit Order Effect, shown in Figure~\ref{fig:prices_vs_wind}.  The effect of increasing the percentage of wind generation in the total energy mix is clearly seen on the SSP, SBP and Spot Price.  All prices decrease with every percentage of wind energy produced.  
There is an observed ``kink'' in all the trendlines around a wind percentage of $30\%$, which is not seen in studies of the foreign markets.  Naively, it would appear that the price increases again above this percentage.  However, we note that this ``kink'' is due to a natural skew in the data: datapoints for which the wind percentage is high (i.e. above $30\%$) occur during the early morning (low SPs) when total demand for electricity (and therefore price) is also low.  If total demand of electricity were to be plotted against the wind percentage, a kink would also be seen at the same position.  The linear fits are detailed in Table~\ref{tab:gradients}, which present both intercepts (x- and y-axis) and the gradients.

% input table of total actual costs and predicted low wind costs
% put here to position it within the paper
\begin{table*}[tb]
\centering
\begin{tabular}{|l|c|c|c|c|c|c|}
\hline
Year & Total Actual & ``No Onshore Wind'' & ``No Offshore Wind'' & ``Low Wind'' & Cost increase & Cost increase \\ 
 & Cost, £bn & Cost, £bn & Cost, £bn & Cost, £bn & £bn & \% \\ \hline

2013  &  36.69  &  38.24  &  38.45  &  39.04  &  2.35  &  6.4 \\
2014  &  29.40  &  30.67  &  31.11  &  31.36  &  1.95  &  6.6 \\ \hline

\end{tabular}
\caption{Annual cost of energy generation: actual cost and predicted volume-weighted total ``no onshore'', ``no offshore'' and ``low wind'' costs per annum.  Due to the reduction of energy prices with significant total wind generation (the Merit-Order Effect from both onshore and offshore wind), the total annual energy bill is heavily reduced by around £2bn per annum. Source: Elexon and APX Group}
\label{tab:low_wind_constraints}
\end{table*}

From the graph, it is reasonable to assume that if wind generation were to be supported in the future, such that higher percentages were reached, the trend would follow the initial lines above the ``kink''. From Table~\ref{tab:gradients}, it can be seen that the price decrease in this range is around \pounds 0.8 - \pounds 1.0 per MWh per extra percent of wind (i.e. $m$ in the table).  Compared to the zero-wind price, a decrease of $\pounds 0.88$ per MWh per \% (i.e. spot price gradient) corresponds to a $1.67\%$ price decrease for every percentage point of wind generation.  This figure broadly agrees with Ketterer who quotes an electricity price drop of $1.46\%$ for a percentage point wind increase in Germany\cite{Ketterer2014270}.

In Figure~\ref{fig:wind_price_contour}, the correlation between onshore and offshore wind generation is compared with the spot price of energy.  The off-centre nature of the contours is due to the imbalance between offshore and onshore components; typically, the UK always produces more electricity from offshore wind and hence the location of the contours follow the population of data points. There is not enough data long the $y=0$ or $x=0$ lines to estimate the Merit-Order Effect for only onshore or offshore wind components.  However, it can be seen that the spot price decreases in roughly concentric circles (centred on zero), which implies that the spot price decreases with total wind generation and does not strongly favour either an offshore or onshore component.  We therefore conclude that it is the total wind percentage that is the important factor when reducing energy prices.  

It is possible to consider the case in which total electricity demand remains the same, but wind generation is minimal.  In this way, comparison between the two cases would present a measure of the cost savings made on energy prices due to wind generation.  It is also possible to estimate the savings cost due to the presence of either an onshore or offshore wind component.  For this, we first calculate the actual cost of electricity in 2013 and 2014.  Since we do not have the cost breakdown for each transaction on the APX market, we instead assume that the price of every unit generated during every settlement period is the spot price.  Since the spot price is a volume-weighted average, this is a reasonable assumption and the total cost should be the correct value.  

%% placed here to position within paper
\begin{figure}[tb!]
\centering
\includegraphics[width=\columnwidth]{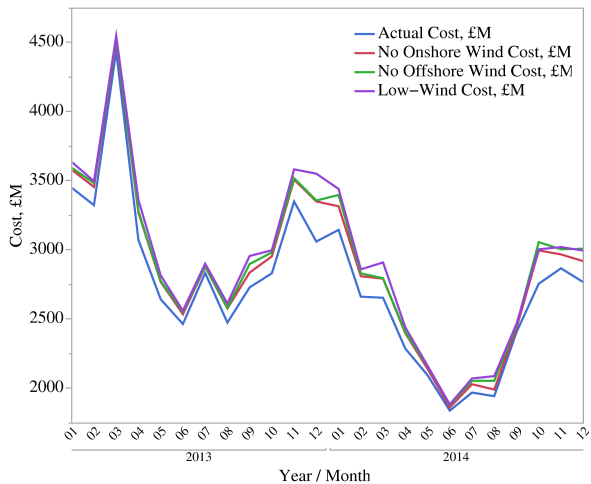}
\caption{The monthly cost of energy for actual generation (blue line) and predicted cost with no onshore wind (red line), no offshore wind (green line) and low-wind generation (purple line). Source: APX group and BMReports} 
\label{fig:Low_wind_costs}
\end{figure}

In order to calculate the savings for each option (``low-wind'', ``no offshore'' and ``no onshore''), we need to simulate the price for each settlement period for each option according to the level of wind generation.  Ideally, we would use the existing dataset to simulate exact fits, but the availability of data is too sparse for this.  Therefore, we bin the data by total wind generation in 5\% categories (0-5\%,5-10\%,10-15\% etc.~with no overlap) and then calculate the volume-weighted mean spot price for each settlement period for each bin.  This leads to a dataset of volume-weighted mean spot prices for each settlement period for each binned category (e.g. volume-weighted mean 0-5\% wind price for each half-hour period).  There are naturally some settlement periods for which there is not enough data to allow for an accurate mean value.  For these few datapoints ($\sim1\%$ of total data), we assume the actual spot price for that settlement period, which should always provide a conservative estimate. 

From this new dataset, it is possible to estimate the spot price for each settlement period for the three options: ``low-wind'', ``no offshore'' and ``no onshore''.  In order to calculate the ``low-wind'' cost, we note that there isn't enough data to consider a true ``zero-wind'' data set.  Therefore, we consider the subset for which wind is less that $5\%$ and utilise this price to calculate the total cost of energy, as if the wind component didn't exist (but assume that the total demand were the same).  For the ``no onshore'' wind option, for each settlement period we assume that the total wind generation percentage is only produced by the offshore wind component (once again assuming the total generation is the same) and use the binned prices calculated above to simulate the cost of energy in this period.  For the ``no offshore'' wind option, we similarly assume that the total wind generation comes from the onshore wind component.  From these costs, the total savings can be simulated for each option.

The simulated monthly costs for each option are shown in Figure~\ref{fig:Low_wind_costs}.  As fully expected, the monthly cost of energy without wind, or with only one component, is higher than the actual cost, due to the Merit Order Effect.  Furthermore, the ``low-wind'', ``no-onshore wind'' and ``no-offshore wind'' costs increase during the winter, when demand is high and actual total wind generation is high.  The total annual savings due to the wind components can be calculated as the difference between the curves.  The statistics are detailed in Table~\ref{tab:low_wind_constraints}.

It can be seen that the cost of providing 717.32 TWh of energy in 2013 was \pounds36.69 billion, but with ``low-wind'' that increased to \pounds39.04 billion.  Therefore, the cost saving due to a wind generation component is estimated to be \pounds2.35 billion in 2013 (6.4\%).  The cost saving in 2014 is \pounds1.95 billion (6.6\%).  

Looking at the complete picture, for the 2013-2014 period, we can see that wind subsidies cost around \pounds3.7 billion, wind curtailment cost \pounds86.2M but the total saving is around \pounds4.3 billion due to the Merit Order Effect.  The net saving is therefore around \pounds514M and (within the errors of our statistical analysis) can be considered at best as a positive effect or conservatively as net zero.

Finally, we can consider the implication of removing a single component of wind (either onshore or offshore).  If there had been no offshore wind, we would have reduce the ROC costs to \pounds1.554 billion over the past two years, but the savings would be reduced to only \pounds840M, resulting in a total outgoing loss of around \pounds797M.  If there had been no onshore wind, the cost of ROCs would have been \pounds2.105 billion and whilst the savings are \pounds1.49 billion, the result still is a loss of around \pounds618M.  This implies that wind generation from both onshore and offshore farms is important to the total economic gain: if either one component is removed, the system would result in a net loss of finance.
% section results (end) 

\section{Conclusions} % (fold)
\label{sec:conclusions}

We have analysed electricity generation datasets within the UK market to study the effect of wind generation and costs (subsidies and curtailment) to the market.  This is of direct relevance in the UK currently, due to the decreasing subsidies for onshore wind.

We have analysed the imbalance markets for relative curtailment costs of wind generation and show that these generators regularly submit negative bids (costs to the system) to curtail generation.  Whilst all generators submit negative bids on occasion, we show that onshore wind generation accounts for the large majority of negative bid cash flow.  Additionally, these bids fall outside of the average bid range for corresponding bids (when compared within each settlement period, as a like-for-like comparison).  The mean negative bid for wind generators is significantly higher than the lost subsidy (around \pounds55/MWh).

We also calculate the savings on the market due to wind generation and the Merit Order Effect.  Compared to the ``zero-wind'' price, we observe a $1.67\%$ price decrease for every percentage point of wind generation, which agrees with analyses of the German market.  (Using the Elexon data to calculate the same effect, we observe a $1.67\%$ price decrease for every percentage point of wind generation).  Due to this effect, we estimate the total cost savings to be around \pounds2 billion per annum, whilst wind subsidies (due to ROCs) cost only around \pounds1.8 billion per annum.  Due to the estimates in this paper (specifically ignoring other subsidies such as FITs which account for around \pounds100M per annum in 2014), we therefore argue that there is an approximate net zero cost to the system due to wind generation and curtailment.

We consider two additional cases, in which there is either no offshore or no onshore wind component.  In both these cases, we find that the Merit-Order Effect is significantly reduced, such that cost savings are also reduced.  The result in both options is a significant final cost to the economy.  We conclude that it is best to have a significant total percentage of wind generation in order to reduce energy prices and both offshore and onshore wind generation are equally important in this effect.
 
% section conclusions (end)

\section{Acknowledgements} % (fold)
\label{sec:acknowledgements}

We thank APX Group for access to their datasets under an academic license.  Thanks also go to Thomas Routier from Elexon, for his considerable help in explaining the imbalance market data to us.  We are hugely grateful to Rt Hon Ed Davey and Good Energy Ltd for their useful comments and suggestions to enhance this work.  This work has been financially supported by EPSRC Grant EP/I032541/1 (``Photovoltaics for Future Societies'').

% section acknowledgements (end)


%% `Elsevier LaTeX' style
%\bibliographystyle{elsarticle-num}

\bibliographystyle{model3-num-names}

\bibliography{wind_pricing.bib}

\end{document}